# Hollow carbon spheres as an efficient dopant for enhancing critical current density of MgB$_2$ based tapes

Zhaoshun Gao, Yanwei Ma[a)], Xianping Zhang, Dongliang Wang, Junhong Wang

Key Laboratory of Applied Superconductivity, Institute of Electrical Engineering, Chinese Academy of Sciences, P. O. Box 2703, Beijing 100190, China

S. Awaji, K. Watanabe

Institute for Materials Research, Tohoku University, Sendai 980-8577, Japan

Boyang Liu

Institute of Materials Science and Engineering, Shanghai Maritime University, Shanghai 201306, China

**Abstract:**

A significant enhancement of critical current density ($J_c$) and $H_{irr}$ in MgB$_2$ tapes has been achieved by the in situ powder-in-tube method utilizing hollow carbon spheres (HCS) as dopants. At 4.2 K, the transport $J_c$ for the 850$^\text{o}$C sintered samples reached $3.1\times10^4$, and $1.4\times10^4$ A/cm$^2$ at 10 and 12 T, respectively, and were better than those of optimal nano-SiC doped tapes. The effects of different sintering temperature on lattice parameters, amount of carbon substitution, microstructure and $J_c$ of doped samples were also investigated. These results demonstrate that HCS is one of the most promising dopants besides nano-carbon and SiC for the enhancement of current capacity for MgB$_2$ in high fields.

---

[a)] Author to whom correspondence should be addressed; E-mail: ywma@mail.iee.ac.cn

# 1. Introduction

Superconducting $MgB_2$ tape has been regarded as a promising candidate for practical applications such as MRI magnet at around 20 K due to its high transition temperature ($T_c$), low material costs and weak-link free grain coupling [1]. The $J_c$ of $MgB_2$ tapes can be easily exceeded the level $10^4$ A/cm$^2$ at 4.2 K and 10 T through chemical doping with carbon containing materials, such as SiC [2-4], C [5,6], $B_4C$ [7], carbon nanotube [8,9], carbohydrate or hydrocarbon [10-13]. The carbon can be incorporated into the $MgB_2$ crystal lattice, and thus $J_c$ and $H_{c2}$ are significantly enhanced due to the enhancement of charge carrier scattering in the two-band $MgB_2$ [14,15]. It should be pointed out that the improvement of superconducting properties is sensitive to morphology of the particles and the synthesis conditions. Doping with these carbon-based materials gives rise to some concerns such as agglomeration of the particles [10], poor reactivity [7], very long milling time [5] or gas release problems [10-13].

In order to overcome these problems, we proposed to use hollow carbon spheres (HCS) as the dopant. TEM images and XRD pattern of the HCS are shown in figure 1. There are some advantages of HCS doping: First, although the diameters of HCS vary from 1 to 8 μm, the nano-scale thickness means that they can be easily broken into nano flakes by ball milling in short time. Second, HCS with a turbostratic graphite structure enables to reduce the friction between the core and the tube walls during deformation process. It is favored for the extent of deformation as well as the core homogeneity of long PIT tapes. In this paper, we have synthesized HCS doped $MgB_2$ tapes and succeeded in significantly improving the transport $J_c$-$H$ properties of $MgB_2$ in magnetic fields up to 14 T.

# 2. Experimental details

The $MgB_2$ composite tapes with HCS doping were prepared by the in situ PIT method. Mg powder (325 mesh, 99.8%), B powder (amorphous, 99.99%) were mixed in a fixed ratio of 1:2 with 1 wt.% HCS powder as additions for the fabrication of tapes. As indicated in Fig.1, the HCS with turbostratic graphite structure and several micrometers in diameter were used in this study. The details of fabricating process and properties for HCS are described elsewhere [16]. The mixtures were put into an agate vial together with agate balls in air atmosphere. The vial and the lid were sealed using an O-ring, then mechanical alloying (MA) was performed in a planetary ball mill for 1 hour. For HCS doped samples, before milling, about 1 ml acetone was dropped to the powder in order to achieve more homogeneity. The packing of the mixed powder into Fe tubes were carried out under a high-purity argon gas atmosphere by using a glove box in order to avoid oxidation and spontaneous combustion of the powders. For the pure samples, the packing of powder were carried out in air. After packing, the tube was rotary swaged and then drawn to wires of 1.75 mm in diameter. The wires were subsequently rolled into tapes. The final size of the tapes was about 3.8 mm in width and about 0.5 mm in thickness. These tapes were heated at 700~850 $^{o}$C for 1 h, and then followed by a furnace cooling to room temperature. The high purity argon gas was allowed to flow into the furnace during the heat-treatment process to reduce the oxidation of the samples.

The phase identification and crystal structure investigation were carried out using Philips X'Pert PRO Diffractometer with Cu K$\alpha$ radiation. Microstructure was studied using a transmission electron microscopy (TEM). DC magnetization measurements were performed with a superconducting quantum interference device (SQUID) magnetometer. Resistivity curves were measured using a current of 100 mA in the high field magnet with Fe sheath, the 10% points on the resistive transition curves were used to define the $H_{irr}$. The transport current $I_c$ at 4.2 K and its magnetic field dependence were evaluated at the High Field Laboratory for Superconducting Materials (HFLSM) at Sendai, by a standard four-probe resistive method, with a criterion of 1 $\mu$V cm$^{-1}$, a magnetic field up to 14 T was applied parallel to the tape surface.

**3. Results and discussion**

Figure 2 shows the results of x-ray diffraction for pure and doped tapes. As can be seen, all samples consisted of a main phase, MgB$_2$, with as mall amount of MgO present. In addition, the Fe$_2$B peaks originating from the reaction layer between iron sheath and core material were also detected. The Mg and B powders were mixed in the stoichiometric ratio, thus probably causing the B losses due to the reaction with the sheath at high temperatures. The reaction layer formation in iron-sheathed tapes is a possible reason for deteriorating transport current [5, 7]. Table I provides an overview of properties for pure and doped $MgB_2$ samples. As shown in table I, the $a$ lattice parameters of all doped samples showed a noticeable decrease compared to the pure samples, while lattice parameter $c$ kept almost unchanged. The shrinkage of $a$ is attributed to the substitution of C for B [5, 17], as the C atom is smaller than the B atom. What is worth noting is that the shrinkage of $a$ lattice clearly larger than that with nano-C doped samples [18]. This result indicates that C coming from HCS seems to substitute easier than nano-C. Furthermore, doping was also confirmed by analysis of full width at half maximum (FWHM) values as shown in Fig. 3. The FWHM of the (002) and (110) peaks for the doped tapes are apparently larger than those of the corresponding peaks for the un-doped one. Broadening in peaks is a strong indication of grain size refinement and lattice distortion, the former can be observed from TEM images of HCS doped $MgB_2$ (Fig. 5). It is worth noting that the influence of doping on (110) peak are apparently larger than (002) peak. On the other hand, FWHM values decreased with increasing sintering temperatures, which may be attributable to the released strains or the improvement in crystallinity [19]. An improvement in crystallinity usually accompanies better grain connectivity of the $MgB_2$ [20], as shown in Table I. As can be seen from Table I, $T_c$ increased with increasing sintering temperature. This $T_c$ increase is a result of improved crystallinity as demonstrated by XRD analyses.

Figure 4 shows the transport $J_c$ at 4.2 K in magnetic fields up to 14 T for HCS doped tapes annealed at 700, 800 and 850 °C, respectively. As a reference, the data of the pure tapes which were sintered at 800 °C are also included in the figure. As can be seen from Fig.4, all doped samples show a drastic enhancement of $J_c$ compared with pure $MgB_2$ tapes. The doped tapes heated at 850 °C reveal the highest $J_c$ compared to all other samples in our experiment: At 4.2 K and in a field of 12 T, the transport $J_c$ reached $1.42\times10^4$ $A/cm^2$, more than 18-fold improvement compared to the pure samples, which have a $J_c$ value about $7.7\times10^2 A/cm^2$. Also, $J_c$ kept $3.1\times10^4$ and $6.0\times10^3$ A/cm$^2$ at 10 and 14 T, respectively. These $J_c$ values of the tapes investigated in this work were better than those of optimal nano-SiC doped tapes [21], and highlight the importance of HCS doping for enhancing the $J_c$ of MgB$_2$ superconductors. We should note that our $J_c$ values are lower than that of carbon doped tapes using harder Monel sheath and high energy ball milling for 50h in Ar [22]. It is expected that further $J_c$ improvements could be obtained by using a harder matrix.

In table I the irreversibility field at 10 K, obtained from the 10% values of their corresponding resistive transitions is presented. As we can see, the $H_{irr}$ properties of the doped MgB$_2$ samples are significantly enhanced. The enhancement of $H_{irr}$ indicates that the intrinsic properties such as band scattering of MgB$_2$ were increased by C substitution into B sites, resulting in higher $H_{c2}$ and hence higher $H_{irr}$. As shown in FWHM results, a distortion of the honeycomb B sheet caused by C substitutional effect is believed to increase the intraband scattering, and thus enhance $H_{c2}$ and $H_{irr}$. From Table 1, electrical resistivity values for pure and HCS doped MgB$_2$ samples at 40 K were 30.9 and 160~210 μΩ cm respectively. The higher ρ values for the doped MgB$_2$ samples indicate that stronger impurity scattering is introduced by the C substitution into B sites.

Figure 5 shows the typical TEM and SEM images for HCS doped tapes. The TEM results reveal that there are a number of impurity phases in the form of nanometer-size inclusions (5–20 nm in size) inside grains in the HCS doped samples [see Figs. 5 (a) 5 (b)]. As the coherence length of MgB$_2$ is about 6–7 nm [23], these 5–20 nm-sized inclusions, with a high density, are ideal flux pinning centres and are responsible for the better $J_c$ performance in doped samples. The selected area diffraction (SAD) pattern, shown in the corner of Fig.5 (a), consists of very well defined ring patterns, suggesting a very fine grain size. The high-resolution TEM images also show very fine MgB$_2$ grains with size about 10-20 nm. Obviously, the fine grain size would create many grain boundaries that may act as effective pinning centres and raise $H_{c2}$ by scattering electrons [24]. The reduction in grain size is probably explained by the inhibition of grain growth by finely dispersed second phases introduced by the doping and milling process [25]. In order to further clarify effects of grain refinement, the SEM images for pure and doped samples sintered at 800 °C were also shown in Fig.5 (e) and (f). From the images, it can be seen that the size of pure $MgB_2$ grains is quite inhomogeneous with some large grains. However, with the addition of HCS, the grain size of $MgB_2$ became small and uniform, which will improve the $J_c$ values.

It is well confirmed now that C substitution into B sites results in an enhancement in $J_c$-$H$ and $H_{irr}$ [2–12]. Due to the special structure of HCS, they can be broken into nano flakes by ball milling and achieve a good mixture with Mg and B. The large distortion of the crystal lattice caused by carbon substitution for B leads to enhanced electron scattering and improvement of $H_{c2}$ and $H_{irr}$. On the other hand, the very fine grain size would create many grain boundaries that may act as effective pinning centres. Therefore, C substitution-induced $H_{irr}$ enhancement as well as the strong flux pinning centers induced by grain boundaries are responsible for the superior $J_c$-$H$ performance of the doped samples. It should be mentioned that the a-lattice parameters of doped samples were around 3.070 Å and did not change even at high sintering temperatures. However, the $J_c$ of sample sintered at 850 °C showed better Jc properties in high fields while slightly reduced $J_c$ of sample sintered at 800 °C. $J_c$ in doped $MgB_2$ tapes may depend on a complex relation between grain connectivity, $H_{c2}$, and flux pinning induced by grain boundaries and precipitates. In our case, the connectivity factor of sample sintered at 850 °C was slightly larger than that at 800 °C. On the other hand, as the sintered temperature increasing, the $MgB_2$ grains become smaller and the distribution of them become more homogeneous (see fig.5). The very fine $MgB_2$ grain with homogeneous distribution will improve the flux pinning density of samples sintered at high temperatures.

## 4. Conclusions

In summary, we have demonstrated that both the transport $J_c$ and $H_{irr}$ of $MgB_2$ tapes are all significantly enhanced by HCS. We observed that sintering the doped wire at low temperature of 700°C, $J_c$ reached $1.2\times10^4$ A/cm$^2$ at 10 T. Further increasing sintering temperatures did not improve amount of C substitution, however, the $J_c$ value was greatly enhanced and reached $3.1\times10^4$ A/cm$^2$ at 10 T for sample sintered at 850°C. This better performance for the 850 °C heated sample can be explained by good connectivity, high flux pinning density induced by very fine grain size.

**Acknowledgments**

The authors thank Haihu Wen, Huan Yang, Liye Xiao and Liangzhen Lin for their help and useful discussion. This work is partially supported by the Beijing Municipal Science and Technology Commission under Grant No. Z07000300700703, National '973' Program (Grant No. 2006CB601004) and National '863' Project (Grant No. 2006AA03Z203).

TABLE I. Some superconducting properties for pure and HCS doped samples

| Sample | Lattice a (Å) | $T_c$ (K) | RRR | $\rho_{40}$ (μΩcm) | $A_F$ (%) | $J_c$ (4.2K, 12 T) | $H_{irr}$ (T) at 10 K |
|---|---|---|---|---|---|---|---|
| Pure | 3.0829(3) | 37.0 | 1.95 | 30.9 | 24.87 | $7.7 \times 10^2$ | 12.4 |
| 700 °C | 3.0706(7) | 30.2 | 1.29 | 208.1 | 12.10 | $5.6 \times 10^3$ | |
| 800 °C | 3.0676(6) | 30.9 | 1.32 | 177.1 | 12.88 | $1.0 \times 10^4$ | 15.8 |
| 850 °C | 3.0692(5) | 31.5 | 1.33 | 168.3 | 13.14 | $1.4 \times 10^4$ | 16.8 |

# Captions

Figure 1  TEM image of HCS showing hollow structure (a), XRD pattern of the HCS (b), HRTEM image of fine structure of the shell near the surface of one HCS is shown in the corner of Fig.1a.

Figure 2  XRD patterns of in situ processed pure and doped tapes. The peaks of $MgB_2$ indexed, while the peaks of MgO and $Fe_2B$ are marked by * and #, respectively.

Figure 3  FWHM values of various diffraction peaks as a function of samples.

Figure 4  $J_c$-$H$ properties of Fe-sheathed pure and doped tapes. The data of nano-SiC and carbon doped tapes are also included for comparison [21, 22].

Figure 5  The TEM images for HCS doped samples sintered at 800 °C (a, c) and 850 °C (b, d). SEM images for the pureand HCS doped samples sintered at 800 °C are also shown in (e) and (f), respectively.

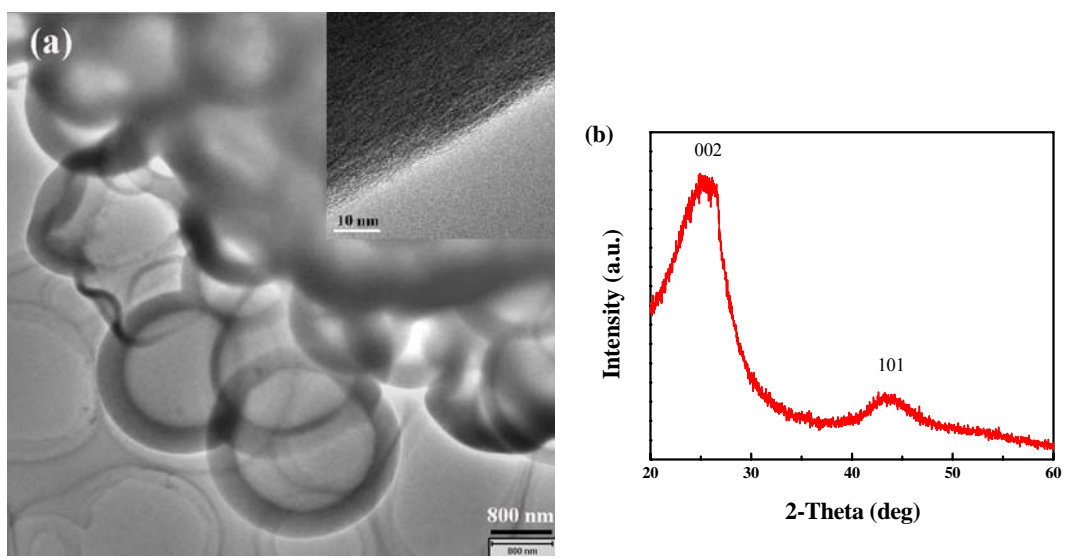

Fig.1 Gao et al.

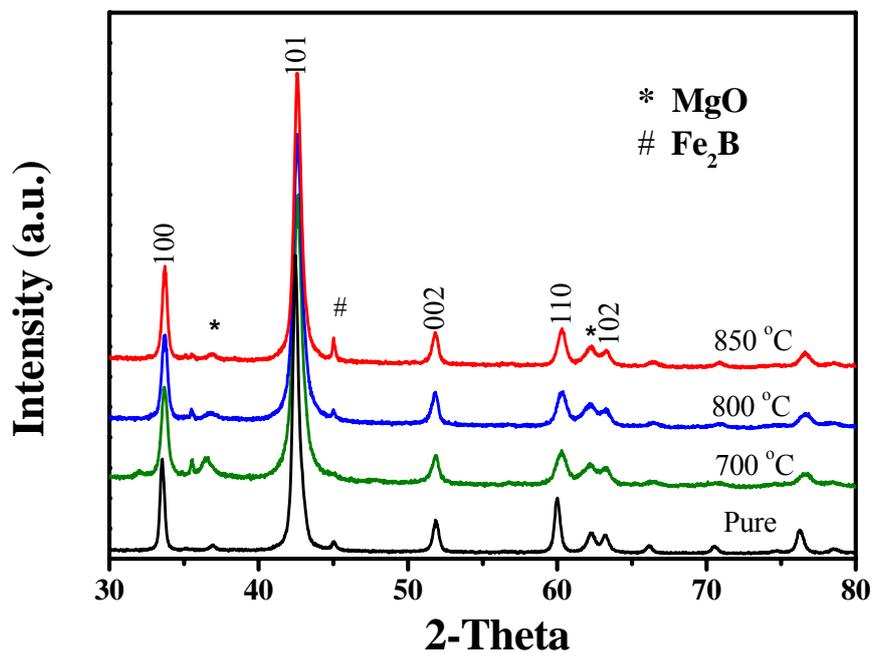

Fig.2 Gao et al.

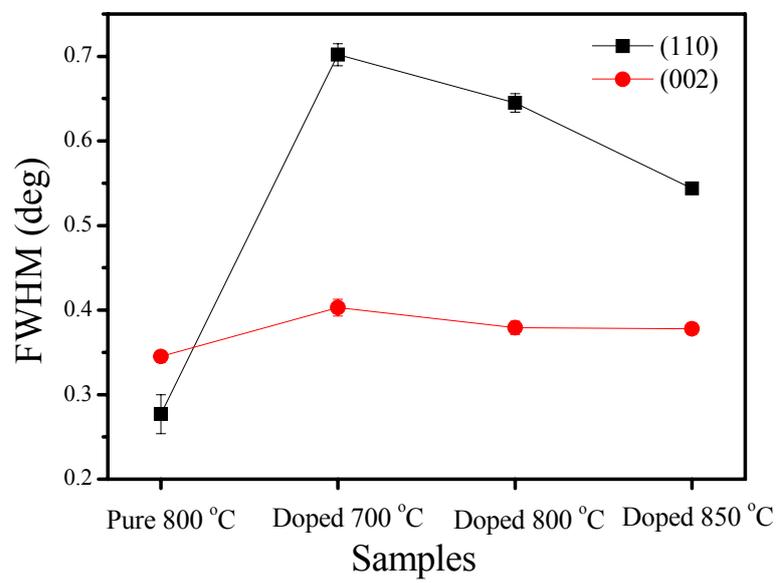

Fig.3 Gao et al.

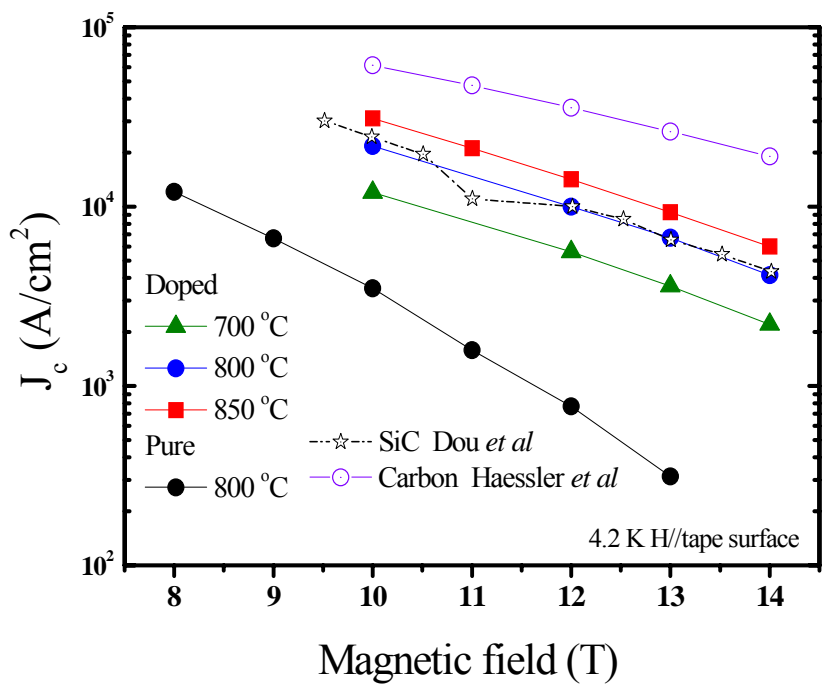

Fig.4 Gao et al.

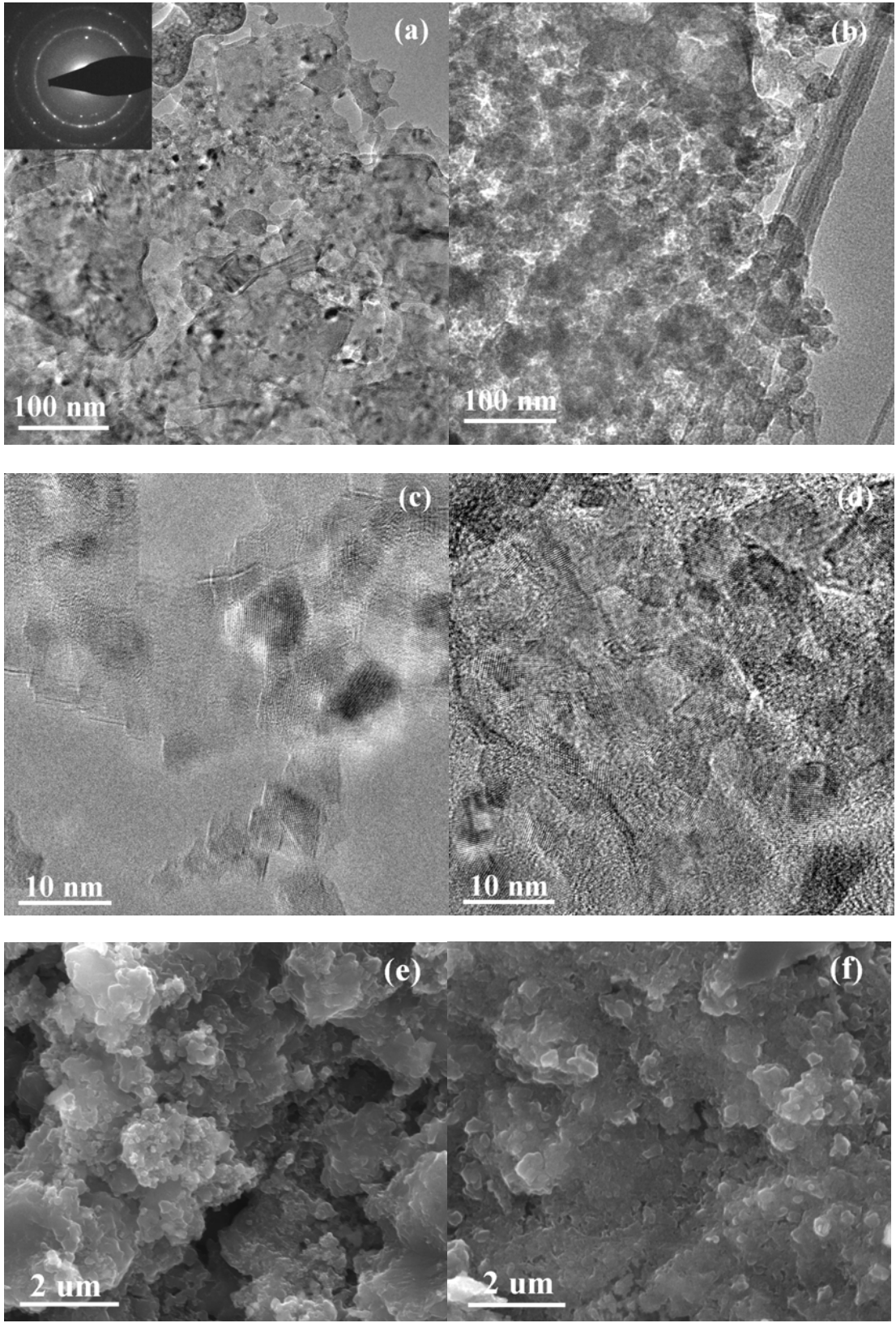

Fig.5 Gao et al.